\documentclass[journal]{IEEEtran}
\usepackage[cmex10]{amsmath}
\usepackage{amsthm}
\usepackage{amssymb}
\usepackage{array}
\usepackage{cite}
\usepackage{graphicx}
\usepackage{hyperref}
\newtheorem{thm}{Theorem}
\newtheorem{conj}[thm]{Conjecture}
\newtheorem{cor}[thm]{Corollary}

\theoremstyle{definition}

\theoremstyle{remark}
\newtheorem{rem}[thm]{Remark}

\begin{document}
\title{New lower bounds to the output entropy of multi-mode quantum Gaussian channels}
\author{Giacomo De Palma
\thanks{Giacomo De Palma is with QMATH, Department of Mathematical Sciences, University of Copenhagen, Universitetsparken 5, 2100 Copenhagen, Denmark}}

\maketitle

\begin{abstract}
We prove that quantum thermal Gaussian input states minimize the output entropy of the multi-mode quantum Gaussian attenuators and amplifiers that are entanglement breaking and of the multi-mode quantum Gaussian phase contravariant channels among all the input states with a given entropy.
This is the first time that this property is proven for a multi-mode channel without restrictions on the input states.
A striking consequence of this result is a new lower bound on the output entropy of all the multi-mode quantum Gaussian attenuators and amplifiers in terms of the input entropy.
We apply this bound to determine new upper bounds to the communication rates in two different scenarios.
The first is classical communication to two receivers with the quantum degraded Gaussian broadcast channel.
The second is the simultaneous classical communication, quantum communication and entanglement generation or the simultaneous public classical communication, private classical communication and quantum key distribution with the Gaussian quantum-limited attenuator.
\end{abstract}


\section{Introduction}
\IEEEPARstart{A}{ttenuation} and noise unavoidably affect electromagnetic communications through wires, optical fibers and free space.
Quantum effects become relevant for low-intensity signals as in the case of satellite communications, where the receiver can be reached by only few photons for each bit of information \cite{chen2012optical}.
Quantum Gaussian channels provide the mathematical model for the attenuation and the noise affecting electromagnetic signals in the quantum regime \cite{chan2006free,braunstein2005quantum,holevo2013quantum,weedbrook2012gaussian,holevo2015gaussian,serafini2017quantum}.

The maximum achievable communication rate of a channel depends on the minimum noise achievable at its output, which is quantified by the output entropy \cite{wilde2017quantum,holevo2013quantum}.
The determination of the maximum rates allowed by quantum mechanics for the communication to two receivers with the quantum degraded Gaussian broadcast channel \cite{guha2007classicalproc,guha2007classical} relies on a minimum output entropy conjecture \cite{guha2008entropy,guha2008capacity} (Conjecture \ref{conj:MOE}).
This fundamental conjecture states that thermal quantum Gaussian input states minimize the output entropy of the quantum Gaussian attenuators, amplifiers and phase contravariant channels among all the input states with a given entropy.
The same conjecture is necessary also to determine the triple trade-off region of the Gaussian quantum-limited attenuator \cite{wilde2012information,wilde2012quantum}.
This region is constituted by all the achievable triples of rates for simultaneous classical communication, quantum communication and entanglement generation or for simultaneous public classical communication, private classical communication and quantum key distribution.
So far, Conjecture \ref{conj:MOE} has been proven only in the special case of one-mode channels \cite{de2015passive,de2016gaussian,de2018pq,de2016gaussiannew,qi2017minimum}.
The best current lower bound to the output entropy of multi-mode quantum Gaussian channels is provided by the quantum Entropy Power Inequality \cite{konig2014entropy,konig2016corrections,de2014generalization,de2015multimode,de2017gaussian,de2018conditional,huber2018conditional,huber2017geometric} (see Theorem \ref{thm:EPI}).
However, this lower bound is strictly lower than the output entropy generated by Gaussian input states, hence it is not sufficient to prove the conjecture (see the review \cite{de2018gaussian} for a complete presentation of the state of the art).

We prove the minimum output entropy conjecture for the multi-mode quantum Gaussian attenuators and amplifiers that are entanglement breaking and for all the multi-mode phase contravariant quantum Gaussian channels (Corollary \ref{cor:EB}).
This is the first time that the minimum output entropy conjecture is proven for a multi-mode channel without restrictions on the input states.
Surprisingly, the implications of this result go beyond the quantum Gaussian channels that are entanglement breaking.
Indeed, combining Corollary \ref{cor:EB} with the quantum integral Stam inequality of Ref. \cite{de2018conditional}, we prove a new lower bound to the output entropy of all the multi-mode quantum Gaussian attenuators and amplifiers (Theorem \ref{thm:main}).
This new lower bound is strictly better than the previous best lower bound provided by the quantum Entropy Power Inequality (see \autoref{fig:epni} for a comparison).

We apply Theorem \ref{thm:main} to determine a new upper bound to the rates for classical communication to two receivers with the quantum degraded Gaussian broadcast channel (Corollary \ref{cor:broadcast}) and a new outer bound to the triple trade-off region of the Gaussian quantum-limited attenuator (Corollary \ref{cor:tradeoff}).
These bounds improve the best previous bounds based on the quantum Entropy Power Inequality (see \autoref{fig:broadcast} and \autoref{fig:tradeoff} for a comparison).

The manuscript is structured as follows.
We present quantum Gaussian channels in \autoref{sec:QGC} and the minimum output entropy conjecture in \autoref{sec:MOE}.
In \autoref{sec:EB} we prove the minimum output entropy conjecture for the quantum Gaussian channels that are entanglement breaking, and in \autoref{sec:main} we prove the new lower bound to the output entropy of the quantum Gaussian attenuators and amplifiers.
We apply this result to prove a new upper bound to the rates for classical communication to two receivers with the quantum degraded Gaussian broadcast channel in \autoref{sec:broadcast} and to prove a new outer bound to the triple trade-off region of the quantum-limited attenuator in \autoref{sec:tradeoff}.
We conclude in \autoref{sec:concl}.

\section{Quantum Gaussian channels}\label{sec:QGC}
A one-mode quantum Gaussian system is the mathematical model for a harmonic oscillator or a mode of the electromagnetic radiation.
The Hilbert space of a one-mode quantum Gaussian system is the irreducible representation of the canonical commutation relation \cite{serafini2017quantum}, \cite[Chapter 12]{holevo2013quantum}
\begin{equation}
\left[\hat{a},\;\hat{a}^\dag\right]=\hat{\mathbb{I}}\;,
\end{equation}
where $\hat{a}$ is the ladder operator.
We define the Hamiltonian
\begin{equation}\label{eq:defH}
\hat{H}=\hat{a}^\dag\hat{a}\;,
\end{equation}
that counts the number of excitations or photons.
The vector annihilated by $\hat{a}$ is the vacuum and is denoted by $|0\rangle$.
A quantum Gaussian state is a quantum state proportional to the exponential of a quadratic polynomial in $\hat{a}$ and $\hat{a}^\dag$.
The most important Gaussian states are the thermal Gaussian states, where the polynomial is proportional to the Hamiltonian \eqref{eq:defH}:
\begin{equation}\label{eq:omega}
\hat{\omega}_E= \frac{1}{\left(E+1\right)}\,\left(\frac{E}{E+1}\right)^{\hat{H}}\;,
\end{equation}
where $E\ge0$ is the average energy:
\begin{equation}
\mathrm{Tr}\left[\hat{H}\,\hat{\omega}_E\right] = E\;.
\end{equation}
We notice that $\hat{\omega}_0 = |0\rangle\langle0|$ is the vacuum state of the system.
The von Neumann entropy of $\hat{\omega}_E$ is
\begin{equation}\label{eq:defg}
S(\hat{\omega}_E) = \left(E+1\right)\ln\left(E+1\right)- E\ln E =: g(E)\;.
\end{equation}
An $n$-mode Gaussian quantum system is the union of $n$ one-mode Gaussian quantum systems, and its Hilbert space is the $n$-th tensor power of the Hilbert space of a one-mode Gaussian quantum system.
Let $\hat{a}_1,\,\ldots,\,\hat{a}_n$ be the ladder operators of the $n$ modes.
The Hamiltonian of the $n$-mode Gaussian quantum system is the sum of the Hamiltonians of each mode:
\begin{equation}
\hat{H} = \sum_{i=1}^n \hat{a}_i^\dag\,\hat{a}_i\;.
\end{equation}

Quantum Gaussian channels are the quantum channels that preserve the set of quantum Gaussian states.
The most important families of quantum Gaussian channels are the beam-splitter, the squeezing, the quantum Gaussian attenuators, the quantum Gaussian amplifiers and the quantum heat semigroup.
The beam-splitter and the squeezing are the quantum counterparts of the classical linear mixing of random variables, and are the main transformations in quantum optics.
Let $A$ and $B$ be one-mode quantum Gaussian systems with ladder operators $\hat{a}$ and $\hat{b}$, respectively.
The \emph{beam-splitter} of transmissivity $0\le\eta\le1$ is implemented by the unitary operator
\begin{equation}\label{eq:defU}
\hat{U}_\eta=\exp\left(\left(\hat{a}^\dag\hat{b}-\hat{b}^\dag\hat{a}\right)\arccos\sqrt{\eta}\right)\;,
\end{equation}
and performs a linear rotation of the ladder operators \cite[Section 1.4.2]{ferraro2005gaussian}:
\begin{align}\label{eq:defUlambda}
\hat{U}_\eta^\dag\,\hat{a}\,\hat{U}_\eta &= \sqrt{\eta}\,\hat{a}+\sqrt{1-\eta}\,\hat{b}\;,\nonumber\\
\hat{U}_\eta^\dag\,\hat{b}\,\hat{U}_\eta &= -\sqrt{1-\eta}\,\hat{a}+\sqrt{\eta}\,\hat{b}\;.
\end{align}
The \emph{squeezing} \cite{barnett2002methods} of parameter $\kappa\ge1$ is implemented by the unitary operator
\begin{equation}\label{eq:defUk}
\hat{U}_\kappa=\exp\left(\left(\hat{a}^\dag\hat{b}^\dag-\hat{a}\,\hat{b}\right)\mathrm{arccosh}\sqrt{\kappa}\right)\;,
\end{equation}
and acts on the ladder operators as
\begin{align}
\hat{U}_\kappa^\dag\,\hat{a}\,\hat{U}_\kappa &= \sqrt{\kappa}\,\hat{a}+\sqrt{\kappa-1}\,\hat{b}^\dag\;,\nonumber\\
\hat{U}_\kappa^\dag\,\hat{b}\,\hat{U}_\kappa &= \sqrt{\kappa-1}\,\hat{a}^\dag+\sqrt{\kappa}\,\hat{b}\;.
\end{align}

The quantum Gaussian attenuators model the attenuation and the noise affecting electromagnetic signals traveling through optical fibers or free space.
The one-mode \emph{quantum Gaussian attenuator} $\mathcal{E}_{\eta,E}$ \cite[case (C) with $k=\sqrt{\lambda}$ and $N_0=E$]{holevo2007one} can be implemented mixing the input state $\hat{\rho}$ with the one-mode thermal Gaussian state $\hat{\omega}_E$ through a beam-splitter of transmissivity $0\le\eta\le1$:
\begin{equation}\label{eq:defE}
\mathcal{E}_{\eta,E}(\hat{\rho}) = \mathrm{Tr}_B\left[\hat{U}_\eta\left(\hat{\rho}\otimes\hat{\omega}_E\right)\hat{U}_\eta^\dag\right]\;.
\end{equation}
If $E=0$ the attenuator is called \emph{quantum-limited}, and we denote
\begin{equation}
\mathcal{E}_{\eta,0} = \mathcal{E}_\eta\;.
\end{equation}
The quantum Gaussian amplifiers model the amplification of electromagnetic signals.
The one-mode \emph{quantum Gaussian amplifier} $\mathcal{A}_{\kappa,E}$ \cite[case (C) with $k=\sqrt{\kappa}$ and $N_0=E$]{holevo2007one} can be implemented performing a squeezing of parameter $\kappa\ge1$ on the input state $\hat{\rho}$ and the one-mode thermal Gaussian state $\hat{\omega}_E$:
\begin{equation}\label{eq:defA}
\mathcal{A}_{\kappa,E}(\hat{\rho}) = \mathrm{Tr}_B\left[\hat{U}_\kappa\left(\hat{\rho}\otimes\hat{\omega}_E\right)\hat{U}_\kappa^\dag\right]\;.
\end{equation}
The one-mode \emph{Gaussian phase contravariant channel} $\tilde{\mathcal{A}}_{\kappa,E}$ \cite[case (D) with $k=\sqrt{\kappa-1}$ and $N_0=E$]{holevo2007one} is the weak complementary of $\mathcal{A}_{\kappa,E}$: for any one-mode quantum state $\hat{\rho}$,
\begin{equation}\label{eq:defAt}
\tilde{\mathcal{A}}_{\kappa,E}(\hat{\rho}) = \mathrm{Tr}_A\left[\hat{U}_\kappa\left(\hat{\rho}\otimes\hat{\omega}_E\right)\hat{U}_\kappa^\dag\right]\;.
\end{equation}

The \emph{displacement operator} $\hat{D}_z$ with $z\in\mathbb{C}$ is the unitary operator that displaces the ladder operators:
\begin{equation}
\hat{D}_z^\dag\,\hat{a}\,\hat{D}_z = \hat{a} + z\,\hat{\mathbb{I}}\;.
\end{equation}
The \emph{quantum Gaussian additive noise channel} $\mathcal{N}_E$ \cite[case ($\text{B}_2$) with $N_c=E$]{holevo2007one} is the quantum Gaussian channel generated by a convex combination of displacement operators with a Gaussian probability measure:
\begin{equation}\label{eq:heat}
\mathcal{N}_E(\hat{\rho}) = \int_{\mathbb{C}}\hat{D}_{\sqrt{E}\,z}\,\hat{\rho}\,\hat{D}_{\sqrt{E}\,z}^\dag\,\mathrm{e}^{-|z|^2}\,\frac{\mathrm{d}z}{\pi}\;,\qquad E>0\;.
\end{equation}

\section{The minimum output entropy conjecture}\label{sec:MOE}

\begin{conj}[minimum output entropy conjecture]\label{conj:MOE}
For any $n\in\mathbb{N}$, quantum Gaussian thermal input states minimize the output entropy of the $n$-mode Gaussian quantum attenuators, amplifiers, phase contravariant channels and additive noise channels among all the input states with a given entropy.
In other words, let $\hat{\rho}$ be a state of an $n$-mode Gaussian quantum system with finite entropy, and let
\begin{equation}
N(\hat{\rho}) = g^{-1}\left(\frac{S(\hat{\rho})}{n}\right)\;,
\end{equation}
where $g$ has been defined in \eqref{eq:defg}, such that $S\left(\hat{\omega}_{N(\hat{\rho})}^{\otimes n}\right) = S(\hat{\rho})$.
Then,
\begin{align}
S\left(\mathcal{E}^{\otimes n}_{\eta,E}(\hat{\rho})\right) &\ge S\left(\mathcal{E}_{\eta,E}^{\otimes n}\left(\hat{\omega}_{N(\hat{\rho})}^{\otimes n}\right)\right)\nonumber\\
& = n\,g\left(\eta\,N(\hat{\rho}) + \left(1-\eta\right)E\right)\;,\nonumber\\
S\left(\mathcal{A}^{\otimes n}_{\kappa,E}(\hat{\rho})\right) &\ge S\left(\mathcal{A}_{\kappa,E}^{\otimes n}\left(\hat{\omega}_{N(\hat{\rho})}^{\otimes n}\right)\right)\nonumber\\
& = n\,g\left(\kappa\,N(\hat{\rho}) + \left(\kappa-1\right)\left(E+1\right)\right)\;,\nonumber\\
S\left(\tilde{\mathcal{A}}^{\otimes n}_{\kappa,E}(\hat{\rho})\right) &\ge S\left(\tilde{\mathcal{A}}_{\kappa,E}^{\otimes n}\left(\hat{\omega}_{N(\hat{\rho})}^{\otimes n}\right)\right)\nonumber\\
& = n\,g\left(\left(\kappa-1\right)\left(N(\hat{\rho})+1\right) + \kappa\,E\right)\;,\nonumber\\
S\left(\mathcal{N}^{\otimes n}_{E}(\hat{\rho})\right) &\ge S\left(\mathcal{N}_{E}^{\otimes n}\left(\hat{\omega}_{N(\hat{\rho})}^{\otimes n}\right)\right)\nonumber\\
& = n\,g\left(N(\hat{\rho}) + E\right)\;.
\end{align}
\end{conj}
\begin{rem}\label{rem:multimode}
Conjecture \ref{conj:MOE} has been proven only in some special cases:
\begin{itemize}
  \item $S(\hat{\rho})=0$, i.e., when $\hat{\rho}$ is pure \cite{giovannetti2015solution,mari2014quantum,holevo2015gaussian};
  \item $n=1$, i.e., one-mode channels (see \cite{de2016gaussian} for the quantum-limited attenuator, \cite{de2016gaussiannew} for all the quantum attenuators, amplifiers and additive noise channels and \cite{qi2017minimum} for the phase contravariant quantum Gaussian channel);
  \item When $\hat{\rho}$ is diagonal in some joint product basis \cite{de2017multimode}.
\end{itemize}
\end{rem}
The current best lower bound to the output entropy of multi-mode quantum Gaussian channels valid for any input state is provided by the quantum Entropy Power Inequality \cite{konig2014entropy,konig2016corrections,de2014generalization,de2015multimode,de2017gaussian,huber2018conditional,huber2017geometric}:
\begin{thm}\label{thm:EPI}
For any $n\in\mathbb{N}$ and any state $\hat{\rho}$ of an $n$-mode Gaussian quantum system with finite average energy,
\begin{align}\label{eq:EPIatt}
S\left(\mathcal{E}^{\otimes n}_{\eta,E}(\hat{\rho})\right) &\ge n\,\ln\left(\eta\exp\frac{S(\hat{\rho})}{n} + \left(1-\eta\right)\exp g(E)\right)\;,\\
\label{eq:EPIampl}
S\left(\mathcal{A}^{\otimes n}_{\kappa,E}(\hat{\rho})\right) &\ge n\,\ln\left(\kappa\exp\frac{S(\hat{\rho})}{n} + \left(\kappa-1\right)\exp g(E)\right)\;,\\
\label{eq:EPIcontr}
S\left(\tilde{\mathcal{A}}^{\otimes n}_{\kappa,E}(\hat{\rho})\right) &\ge n\,\ln\left(\left(\kappa-1\right)\exp\frac{S(\hat{\rho})}{n} + \kappa\,\exp g(E)\right)\\
\label{eq:EPIadd}
S\left(\mathcal{N}^{\otimes n}_{E}(\hat{\rho})\right) &\ge n\,\ln\left(\exp\frac{S(\hat{\rho})}{n} + \mathrm{e}\,E\right)\;.
\end{align}
\begin{proof}
The claim \eqref{eq:EPIatt} follows from the quantum Entropy Power Inequality for the beam-splitter \cite[Eq. (5)]{de2014generalization} and the representation \eqref{eq:defE} for the quantum Gaussian attenuator.
The claim \eqref{eq:EPIampl} and \eqref{eq:EPIcontr} follow from the quantum Entropy Power Inequality for the squeezing \cite[Eq. (7)]{de2014generalization} and the representations \eqref{eq:defA} and \eqref{eq:defAt} for the quantum phase contravariant Gaussian channel.
The claim \eqref{eq:EPIadd} follows from \cite[Theorem 3]{huber2017geometric}.
\end{proof}
\end{thm}

\section{Gaussian states minimize the output entropy of entanglement breaking quantum Gaussian channels}\label{sec:EB}
In this Section, we prove Conjecture \ref{conj:MOE} for the phase contravariant quantum Gaussian channels and for the quantum Gaussian attenuators and amplifiers that are entanglement breaking.
This result is a corollary of the following.
\begin{thm}\label{thm:EB}
Let $A$ and $B$ be quantum systems with Hilbert spaces $\mathcal{H}_A$ and $\mathcal{H}_B$, and let $\Phi:A\to B$ be an entanglement breaking quantum channel such that for any quantum state $\hat{\rho}$ on $\mathcal{H}_A$
\begin{equation}\label{eq:deff}
S(\Phi(\hat{\rho})) \ge f(S(\hat{\rho}))\;,
\end{equation}
with $f:[0,\infty)\to[0,\infty)$ increasing and convex.
Then, for any $n\in\mathbb{N}$ and any quantum state $\hat{\rho}$ on $\mathcal{H}_A^{\otimes n}$,
\begin{equation}\label{eq:ebclaim}
S\left(\Phi^{\otimes n}(\hat{\rho})\right) \ge n\,f\left(\frac{S(\hat{\rho})}{n}\right)\;.
\end{equation}
\begin{proof}
We prove the claim by induction on $n$.
The claim is true for $n=1$.
Let us then assume \eqref{eq:ebclaim} for a given $n$.
Let $\hat{\rho}_{A_1\ldots A_{n+1}}$ be a quantum state on $\mathcal{H}_A^{\otimes\left(n+1\right)}$, and let
\begin{equation}
\hat{\rho}_{B_1\ldots B_{n+1}} = \Phi^{\otimes\left(n+1\right)}(\hat{\rho}_{A_1\ldots A_{n+1}})\;.
\end{equation}
Since $\Phi$ is entanglement breaking, it admits a representation as a measure-prepare channel \cite{kholevo2005notion}, i.e., there exist a complete separable metric space $X$, a quantum-classical channel $\Phi_1$ that maps quantum states on $A$ to Borel probability measures on $X$ and a classical-quantum channel $\Phi_2$ that maps Borel probability measures on $X$ to quantum states on $B$ such that
\begin{equation}
\Phi = \Phi_2\circ\Phi_1\;.
\end{equation}
We define the probability measure on $X$ taking values on quantum states on $\mathcal{H}_A^{\otimes n}$
\begin{equation}
\hat{\rho}_{A_1\ldots A_n X} = \left(\mathbb{I}_{A_1\ldots A_n}\otimes \Phi_1\right)(\hat{\rho}_{A_1\ldots A_{n+1}})\;,
\end{equation}
and the probability measure on $X$ taking values on quantum states on $\mathcal{H}_B^{\otimes n}$
\begin{equation}
\hat{\rho}_{B_1\ldots B_n X} = \left(\Phi^{\otimes n}\otimes \mathbb{I}_X\right)(\hat{\rho}_{A_1\ldots A_nX})\;,
\end{equation}
such that
\begin{equation}
\hat{\rho}_{B_1\ldots B_{n+1}} = \left(\mathbb{I}_{B_1\ldots B_n}\otimes\Phi_2\right)(\hat{\rho}_{B_1\ldots B_n X})\;.
\end{equation}
We have
\begin{align}\label{eq:cineq}
&S(B_1\ldots B_n|X) = \int_X S(B_1\ldots B_n|X=x)\,\mathrm{d}\rho_X(x)\nonumber\\
&\ge n\int_X f\left(\frac{S(A_1\ldots A_n|X=x)}{n}\right)\mathrm{d}\rho_X(x)\nonumber\\
&\ge n\,f\left(\frac{1}{n}\int_XS(A_1\ldots A_n|X=x)\,\mathrm{d}\rho_X(x)\right)\nonumber\\
&= n\,f\left(\frac{S(A_1\ldots A_n|X)}{n}\right)\;,
\end{align}
where we have used the inductive hypothesis \eqref{eq:ebclaim} and Jensen's inequality applied to the convex function $f$.
We then have
\begin{align}
&S(B_1\ldots B_{n+1}) \overset{\text{(a)}}{=} S(B_{n+1}) + S(B_1\ldots B_n|B_{n+1})\nonumber\\
&\overset{\text{(b)}}{\ge} S(B_{n+1}) + S(B_1\ldots B_n|X_{n+1})\nonumber\\
&\overset{\text{(c)}}{\ge} f(S(A_{n+1})) + n\,f\left(\frac{S(A_1\ldots A_n|X_{n+1})}{n}\right)\nonumber\\
&\overset{\text{(d)}}{\ge} f(S(A_{n+1})) + n\,f\left(\frac{S(A_1\ldots A_n|A_{n+1})}{n}\right)\nonumber\\
&\overset{\text{(e)}}{\ge} \left(n+1\right)f\left(\frac{S(A_{n+1}) + S(A_1\ldots A_n|A_{n+1})}{n+1}\right)\nonumber\\
&\overset{\text{(f)}}{=}\left(n+1\right)f\left(\frac{S(A_1\ldots A_{n+1})}{n+1}\right)\;.
\end{align}
(a) follows from the chain rule for the entropy; (b) follows from the data processing inequality for the channel $\Phi_2$; (c) follows from the hypothesis \eqref{eq:deff} and from \eqref{eq:cineq}; (d) follows from the data processing inequality for the channel $\Phi_1$ (we recall that $f$ is increasing); (e) follows from Jensen's inequality applied to the convex function $f$; (f) follows from the chain rule for the entropy.
We have then proven that the claim \eqref{eq:ebclaim} for $n$ implies the claim \eqref{eq:ebclaim} for $n+1$, and by induction the claim is true for any $n$.
\end{proof}
\end{thm}
The following Corollary \ref{cor:EB} proves Conjecture \ref{conj:MOE} for all the channels that are entanglement breaking.
This is the first time that Conjecture \ref{conj:MOE} is proven for multi-mode channels without restrictions on the input states.
\begin{cor}[minimum output entropy conjecture for entanglement breaking channels]\label{cor:EB}
Conjecture \ref{conj:MOE} holds for:
\begin{itemize}
  \item Any quantum Gaussian attenuator $\mathcal{E}_{\eta,E}$ with $E\ge\tfrac{\eta}{1-\eta}$;
  \item Any quantum Gaussian amplifier $\mathcal{A}_{\kappa,E}$ with $E\ge\tfrac{1}{\kappa-1}$;
  \item Any quantum Gaussian phase contravariant channel $\tilde{A}_{\kappa,E}$;
  \item Any quantum Gaussian additive noise channel $\mathcal{N}_E$ with $E\ge1$.
\end{itemize}
\begin{proof}
Conjecture \ref{conj:MOE} holds for $n=1$.
From \cite[Sec. 12.6.2]{holevo2013quantum}, the conditions $E\ge\tfrac{\eta}{1-\eta}$, $E\ge\tfrac{1}{\kappa-1}$ and $E\ge1$ imply that $\mathcal{E}_{\eta,E}$, $\mathcal{A}_{\kappa,E}$ and $\mathcal{N}_E$ are entanglement breaking, respectively, and $\tilde{\mathcal{A}}_{\kappa,E}$ is entanglement breaking for any $E\ge0$.
From \cite[Lemma 15]{de2017multimode}, the functions
\begin{align}
x&\mapsto g\left(\eta\,g^{-1}(x) + \left(1-\eta\right)E\right)\;,\nonumber\\
x&\mapsto g\left(\kappa\,g^{-1}(x) + \left(\kappa-1\right)\left(E+1\right)\right)\;,\nonumber\\
x&\mapsto g\left(\left(\kappa-1\right)\left(g^{-1}(x)+1\right) + \kappa\,E\right)\;,\nonumber\\
x&\mapsto g\left(g^{-1}(x) + E\right)
\end{align}
are increasing and convex for any $0\le\eta\le1$, $\kappa\ge1$ and $E\ge0$.
The claim then follows from Theorem \ref{thm:EB}.
\end{proof}
\end{cor}

\section{The new lower bound to the output entropy of quantum Gaussian channels}\label{sec:main}
\begin{figure}
\includegraphics[width=\columnwidth]{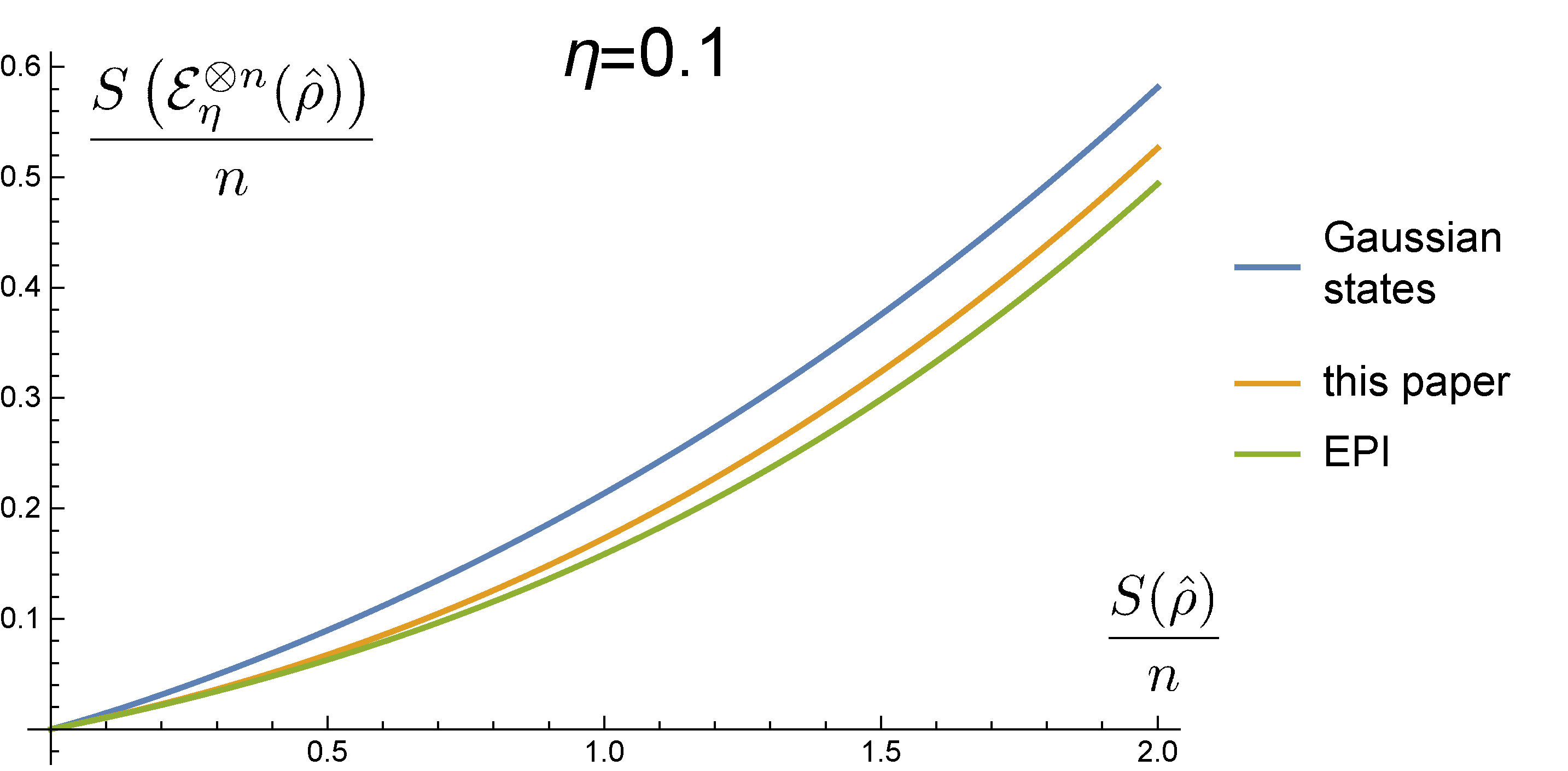}
\includegraphics[width=\columnwidth]{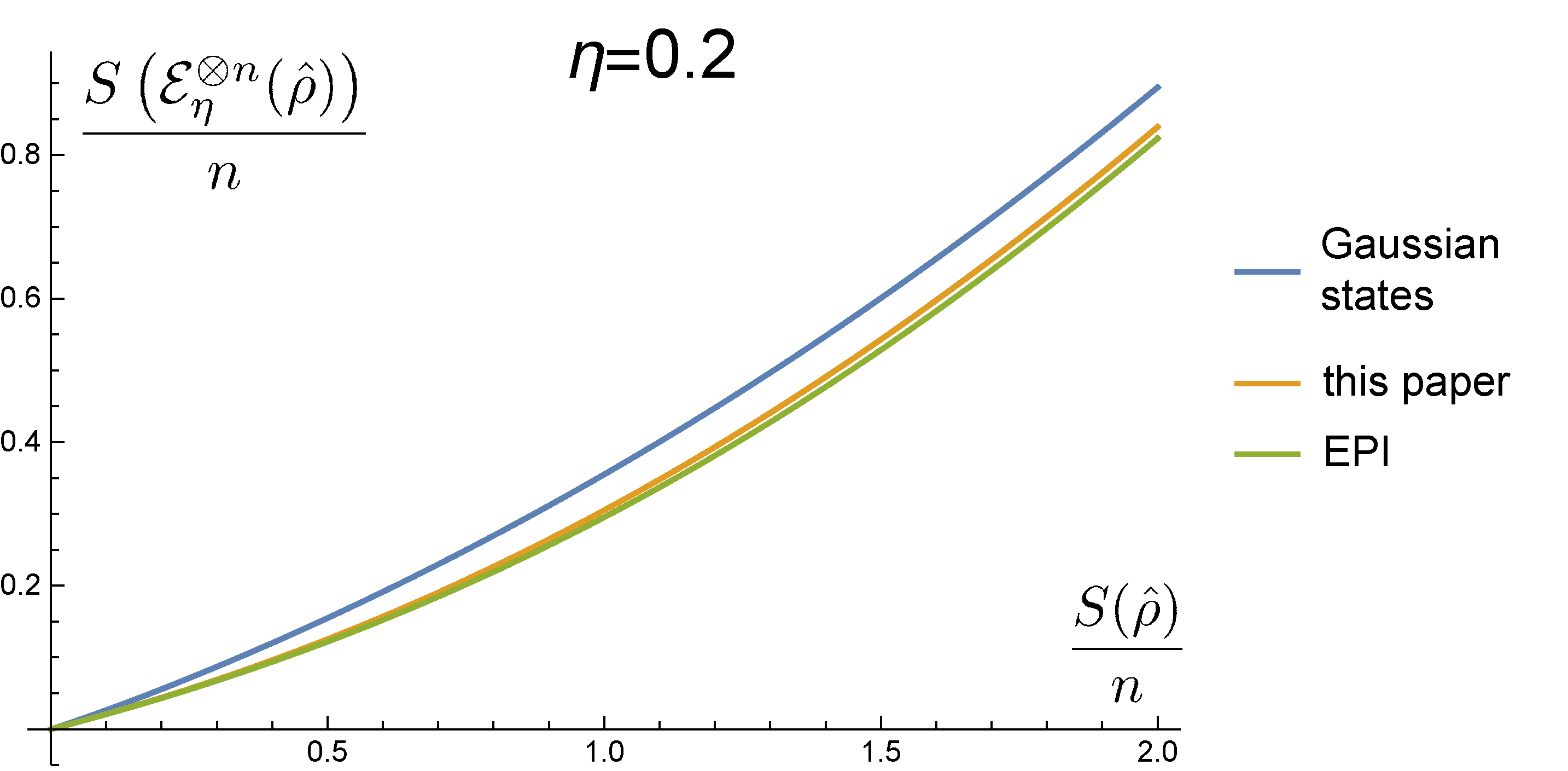}
\caption{Output entropy of the Gaussian quantum-limited attenuator with attenuation parameter $\eta = 0.1,\,0.2$ as a function of the input entropy.
The plot compares the output entropy achieved by thermal Gaussian input states with the lower bounds provided by Theorem \ref{thm:main} and by the quantum Entropy Power Inequality.}\label{fig:epni}
\end{figure}

A striking consequence of Corollary \ref{cor:EB} is the following improved lower bound for the output entropy of the multi-mode quantum Gaussian channels that are not entanglement breaking.
We compare in \autoref{fig:epni} this bound with the previous best bound provided by the quantum Entropy Power Inequality and with the output entropy achieved by quantum thermal Gaussian input states.
\begin{thm}\label{thm:main}
For any $n\in\mathbb{N}$ and any state $\hat{\rho}$ of an $n$-mode Gaussian quantum system with finite average energy,
\begin{align}
\frac{S\left(\mathcal{E}^{\otimes n}_{\eta,E}(\hat{\rho})\right)}{n} &\ge g\left(\eta\,g^{-1}\left(\frac{S(\hat{\rho})}{n} + g\left(\tfrac{\eta}{1-\eta}\right) - g(E)\right) + \eta\right) \nonumber\\
&\phantom{\ge} + g(E) - g\left(\tfrac{\eta}{1-\eta}\right)\qquad\forall\;0\le E\le \tfrac{\eta}{1-\eta}\;,\nonumber\\
\frac{S\left(\mathcal{A}^{\otimes n}_{\kappa,E}(\hat{\rho})\right)}{n} &\ge g\left(\kappa g^{-1}\left(\frac{S(\hat{\rho})}{n} + g\left(\tfrac{1}{\kappa-1}\right) - g(E)\right) + \kappa\right)\nonumber\\
&\phantom{\ge} + g(E) - g\left(\tfrac{1}{\kappa-1}\right)\qquad\forall\;0\le E\le\tfrac{1}{\kappa-1}\;,\nonumber\\
\frac{S\left(\mathcal{N}^{\otimes n}_{E}(\hat{\rho})\right)}{n} &\ge g\left(g^{-1}\left(\frac{S(\hat{\rho})}{n}  - \ln E\right) + 1\right) + \ln E\nonumber\\
& \forall\;0\le E\le 1\;.
\end{align}
\begin{proof}
{\bf Quantum Gaussian attenuators.}
We fix $0\le\lambda\le 1$ and define for any $t\ge0$
\begin{align}
\hat{\rho}(t) &= \mathcal{N}^{\otimes n}_\frac{\lambda\,t}{\eta}(\hat{\rho})\;,\qquad E(t) = E + \frac{1-\lambda}{1-\eta}\,t\;,\nonumber\\
\phi(t) &= S\left(\mathcal{E}^{\otimes n}_{\eta,E(t)}(\hat{\rho}(t))\right)\nonumber\\
& \phantom{=}- \lambda\,S(\hat{\rho}(t)) - \left(1-\lambda\right)S\left({\hat{\omega}(E(t))}^{\otimes n}\right)\;.
\end{align}
From \cite[Eq. (113)]{de2018conditional} we have $\phi(t)\le\phi(0)$, hence
\begin{align}\label{eq:phi}
\frac{S\left(\mathcal{E}^{\otimes n}_{\eta,E}(\hat{\rho})\right)}{n} &\ge \frac{S\left(\mathcal{E}^{\otimes n}_{\eta,E(t)}(\hat{\rho}(t))\right)}{n} -\lambda\,\frac{S(\hat{\rho}(t)) - S(\hat{\rho})}{n}\nonumber\\
&\phantom{\ge} - \left(1-\lambda\right)\left(g(E(t)) - g(E)\right)\;.
\end{align}
We set
\begin{equation}
t = t^* = \frac{\eta - \left(1-\eta\right)E}{1-\lambda}\;,
\end{equation}
such that $E(t^*) = \tfrac{\eta}{1-\eta}$ and the channel $\mathcal{E}_{\eta,E(t^*)}$ is entanglement breaking.
Then, putting together \eqref{eq:phi} and Corollary \ref{cor:EB} we get
\begin{align}\label{eq:EBineq}
\frac{S\left(\mathcal{E}^{\otimes n}_{\eta,E}(\hat{\rho})\right)}{n} &\ge f\left(\frac{S(\hat{\rho}(t^*))}{n}\right) -\lambda\,\frac{S(\hat{\rho}(t^*)) - S(\hat{\rho})}{n}\nonumber\\
&\phantom{\ge} - \left(1-\lambda\right)\left(g\left(\tfrac{\eta}{1-\eta}\right) - g(E)\right)\;,
\end{align}
where for any $x\ge0$
\begin{equation}
f(x) = g(\eta\,g^{-1}(x) + \eta)\;.
\end{equation}
Let
\begin{equation}
S_0 = \frac{S(\hat{\rho})}{n} + g\left(\tfrac{\eta}{1-\eta}\right) - g(E)\;.
\end{equation}
From \cite[Lemma 15]{de2017multimode}, $f$ is convex, hence
\begin{equation}\label{eq:Jensen}
f\left(\frac{S(\hat{\rho}(t^*))}{n}\right) \ge f(S_0) + \left(\frac{S(\hat{\rho}(t^*))}{n} - S_0\right)f'(S_0)\;.
\end{equation}
Finally, we set $\lambda = f'(S_0)$ and get from \eqref{eq:EBineq} and \eqref{eq:Jensen}
\begin{equation}
\frac{S\left(\mathcal{E}^{\otimes n}_{\eta,E}(\hat{\rho})\right)}{n} \ge f(S_0) + g(E) - g\left(\tfrac{\eta}{1-\eta}\right)\;,
\end{equation}
and the claim follows.

{\bf Quantum Gaussian amplifiers.}
The proof for the quantum Gaussian amplifiers is analogous to the proof for the quantum Gaussian attenuators.
We fix $0\le\lambda\le 1$ and define for any $t\ge0$
\begin{align}
\hat{\rho}(t) &= \mathcal{N}^{\otimes n}_\frac{\lambda\,t}{\kappa}(\hat{\rho})\;,\qquad E(t) = E + \frac{1-\lambda}{\kappa-1}\,t\;,\nonumber\\
\phi(t) &= S\left(\mathcal{A}^{\otimes n}_{\kappa,E(t)}(\hat{\rho}(t))\right)\nonumber\\
&\phantom{=} - \lambda\,S(\hat{\rho}(t)) - \left(1-\lambda\right)S\left({\hat{\omega}(E(t))}^{\otimes n}\right)\;.
\end{align}
From \cite[Eq. (113)]{de2018conditional} we have $\phi(t)\le\phi(0)$, hence
\begin{align}\label{eq:phiA}
\frac{S\left(\mathcal{A}^{\otimes n}_{\kappa,E}(\hat{\rho})\right)}{n} &\ge \frac{S\left(\mathcal{A}^{\otimes n}_{\kappa,E(t)}(\hat{\rho}(t))\right)}{n} -\lambda\,\frac{S(\hat{\rho}(t)) - S(\hat{\rho})}{n}\nonumber\\
&\phantom{\ge} - \left(1-\lambda\right)\left(g(E(t)) - g(E)\right)\;.
\end{align}
We set
\begin{equation}
t = t^* = \frac{1 - \left(\kappa-1\right)E}{1-\lambda}\;,
\end{equation}
such that $E(t^*) = \tfrac{1}{\kappa-1}$ and the channel $\mathcal{A}_{\kappa,E(t^*)}$ is entanglement breaking.
Then, putting together \eqref{eq:phiA} and Corollary \ref{cor:EB} we get
\begin{align}\label{eq:EBineqA}
\frac{S\left(\mathcal{A}^{\otimes n}_{\kappa,E}(\hat{\rho})\right)}{n} &\ge f\left(\frac{S(\hat{\rho}(t^*))}{n}\right) -\lambda\,\frac{S(\hat{\rho}(t^*)) - S(\hat{\rho})}{n}\nonumber\\
&\phantom{\ge} - \left(1-\lambda\right)\left(g\left(\tfrac{1}{\kappa-1}\right) - g(E)\right)\;,
\end{align}
where for any $x\ge0$
\begin{equation}
f(x) = g(\kappa\,g^{-1}(x) + \kappa)\;.
\end{equation}
Let
\begin{equation}
S_0 = \frac{S(\hat{\rho})}{n} + g\left(\tfrac{1}{\kappa-1}\right) - g(E)\;.
\end{equation}
From \cite[Lemma 15]{de2017multimode}, $f$ is convex, hence
\begin{equation}\label{eq:JensenA}
f\left(\frac{S(\hat{\rho}(t^*))}{n}\right) \ge f(S_0) + \left(\frac{S(\hat{\rho}(t^*))}{n} - S_0\right)f'(S_0)\;.
\end{equation}
Finally, we set $\lambda = f'(S_0)$ and get from \eqref{eq:EBineqA} and \eqref{eq:JensenA}
\begin{equation}
\frac{S\left(\mathcal{A}^{\otimes n}_{\kappa,E}(\hat{\rho})\right)}{n} \ge f(S_0) + g(E) - g\left(\tfrac{1}{\kappa-1}\right)\;,
\end{equation}
and the claim follows.

{\bf Quantum Gaussian additive noise channels.}
We fix $0\le\lambda\le 1$ and define for any $t\ge0$
\begin{align}
\hat{\rho}(t) &= \mathcal{N}^{\otimes n}_{\lambda\,t}(\hat{\rho})\;,\qquad E(t) = E + \left(1-\lambda\right)t\;,\nonumber\\
\phi(t) &= S\left(\mathcal{N}^{\otimes n}_{E(t)}(\hat{\rho}(t))\right) - \lambda\,S(\hat{\rho}(t)) - n\left(1-\lambda\right)\ln E(t)\;.
\end{align}
From the proof of Theorem 5 of Ref. \cite{huber2018conditional} we have $\phi(t)\le\phi(0)$, hence
\begin{align}\label{eq:phiN}
\frac{S\left(\mathcal{N}^{\otimes n}_E(\hat{\rho})\right)}{n} &\ge \frac{S\left(\mathcal{N}^{\otimes n}_{E(t)}(\hat{\rho}(t))\right)}{n} \nonumber\\
&\phantom{\ge} -\lambda\,\frac{S(\hat{\rho}(t)) - S(\hat{\rho})}{n} - \left(1-\lambda\right)\ln\frac{E(t)}{E}\;.
\end{align}
We set
\begin{equation}
t = t^* = \frac{1 - E}{1-\lambda}\;,
\end{equation}
such that $E(t^*) = 1$ and the channel $\mathcal{N}_{E(t^*)}$ is entanglement breaking.
Then, putting together \eqref{eq:phiN} and Corollary \ref{cor:EB} we get
\begin{align}\label{eq:EBineqN}
\frac{S\left(\mathcal{N}^{\otimes n}_E(\hat{\rho})\right)}{n} &\ge f\left(\frac{S(\hat{\rho}(t^*))}{n}\right) \nonumber\\
&\phantom{\ge} -\lambda\,\frac{S(\hat{\rho}(t^*)) - S(\hat{\rho})}{n} + \left(1-\lambda\right)\ln E\;,
\end{align}
where for any $x\ge0$
\begin{equation}
f(x) = g(g^{-1}(x) + 1)\;.
\end{equation}
Let
\begin{equation}
S_0 = \frac{S(\hat{\rho})}{n} - \ln E\;.
\end{equation}
From \cite[Lemma 15]{de2017multimode}, $f$ is convex, hence
\begin{equation}\label{eq:JensenN}
f\left(\frac{S(\hat{\rho}(t^*))}{n}\right) \ge f(S_0) + \left(\frac{S(\hat{\rho}(t^*))}{n} - S_0\right)f'(S_0)\;.
\end{equation}
Finally, we set $\lambda = f'(S_0)$ and get from \eqref{eq:EBineqN} and \eqref{eq:JensenN}
\begin{equation}
\frac{S\left(\mathcal{N}^{\otimes n}_E(\hat{\rho})\right)}{n} \ge f(S_0) + \ln E\;,
\end{equation}
and the claim follows.
\end{proof}
\end{thm}
\begin{rem}
Since states with infinite average energy are unphysical, for all practical purposes the hypothesis of finite average energy in Theorem \ref{thm:main} is not restrictive.
\end{rem}

\section{Bound to the capacity region of the quantum degraded Gaussian broadcast channel}\label{sec:broadcast}
Let $A$, $B$, $A'$, $B'$ be one-mode Gaussian quantum systems.
The quantum degraded Gaussian broadcast channel \cite{guha2007classicalproc,guha2007classical} maps a state $\hat{\rho}_A$ of $A$ to a state $\hat{\rho}_{A'B'}$ of the joint quantum system $A'B'$ with
\begin{equation}\label{eq:defbr}
\hat{\rho}_{A'B'}=\hat{U}_\eta\left(\hat{\rho}_A\otimes|0\rangle_B\langle0|\right)\hat{U}_\eta^\dag\;,
\end{equation}
where $\hat{U}_\eta$ is the unitary operator defined in \eqref{eq:defU} and $\frac{1}{2}\leq\eta\leq1$.
The channel can be understood as follows.
$A$ encodes the information into the state of the electromagnetic radiation $\hat{\rho}_A$, and sends it through a beam-splitter of transmissivity $\eta$.
$A'$ and $B'$ receive the transmitted and the reflected part of the signal, respectively, whose joint state is $\hat{\rho}_{A'B'}$.
This channel is called degraded since the state received by $B'$ can be obtained applying a quantum-limited attenuator to the state received by $A'$ \cite{guha2007classical}:
\begin{equation}\label{eq:degr}
\hat{\rho}_{B'} = \mathcal{E}_\frac{1-\eta}{\eta}(\hat{\rho}_{A'})\;.
\end{equation}

The simplest communication strategy is time sharing, which consists in communicating only with $A'$ for a fraction of the time and only with $B'$ for the remaining fraction of the time.
Superposition coding \cite{yard2011quantum,guha2007classical,savov2015classical} is a more sophisticated strategy that achieves higher rates communicating with $A'$ and $B'$ simultaneously.
Let $E>0$ be the maximum average energy per mode of the input states.
Superposition coding allows to achieve with the quantum degraded Gaussian broadcast channel \eqref{eq:defbr} any rate pair $(R_{A'},\,R_{B'})$ satisfying \cite[Sec. IV]{guha2007classical}
\begin{equation}\label{eq:R}
R_{A'}\ge0\;,\quad 0\le R_{B'} \le g((1-\eta)E) - g\left(\tfrac{1-\eta}{\eta}\,g^{-1}(R_{A'})\right)\;.
\end{equation}
Assuming Conjecture \ref{conj:MOE} for the quantum-limited attenuator, the capacity region of the quantum degraded Gaussian broadcast channel coincides with the region identified by \eqref{eq:R} \cite{guha2007classical}, i.e., any achievable rate pair satisfies \eqref{eq:R}.

\begin{figure}
  \includegraphics[width=\columnwidth]{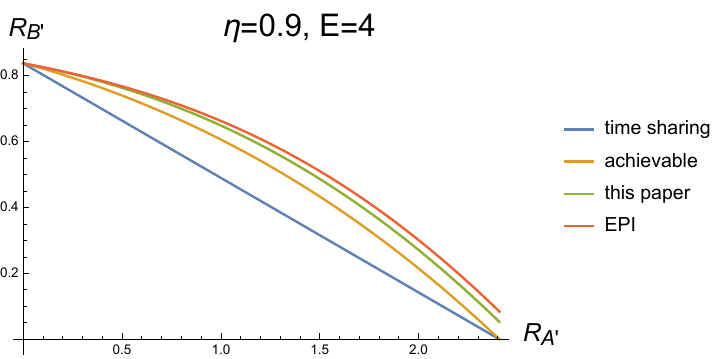}
  \caption{Capacity region in nats for the communication to two receivers with the quantum degraded Gaussian broadcast channel with parameter $\eta=0.9$ and input states with at most $E=4$ average photons per mode.
  The plot compares the regions achievable with time sharing and with the superposition coding \eqref{eq:R} to the outer bounds provided by the quantum Entropy Power Inequality and by Corollary \ref{cor:broadcast}.
  All the rates are expressed in nats per channel use.}\label{fig:broadcast}
\end{figure}

Despite Conjecture \ref{conj:MOE} still lacks a proof, the known lower bounds to the output entropy of the multi-mode quantum-limited attenuators still imply bounds to the capacity region of the quantum degraded Gaussian broadcast channel.
The first of these bounds has been determined from the quantum Entropy Power Inequality \cite{de2014generalization}.
The following Theorem \ref{thm:broadf} shows that any lower bound to the output entropy of the multi-mode quantum-limited attenuators in terms of the input entropy implies a bound to the capacity region of the quantum degraded Gaussian broadcast channel.
We then combine Theorem \ref{thm:broadf} with Theorem \ref{thm:main} to obtain a new bound to this capacity region.
\begin{thm}\label{thm:broadf}
Let us suppose that for any $n\in\mathbb{N}$, any $0\le\lambda\le1$ and any input state $\hat{\rho}$ of an $n$-mode Gaussian quantum system with finite average energy
\begin{equation}\label{eq:f}
S\left(\mathcal{E}^{\otimes n}_\lambda(\hat{\rho})\right) \ge n\,f_\lambda\left(\frac{S(\hat{\rho})}{n}\right)\;,
\end{equation}
where the function $f_\lambda$ is increasing and convex.
Then, any achievable rate pair $(R_{A'},\,R_{B'})$ for the quantum degraded Gaussian broadcast channel satisfies
\begin{equation}\label{eq:Rf}
R_{A'}\ge0\;,\qquad 0\le R_{B'} \le g\left(\left(1-\eta\right)E\right) - f_\frac{1-\eta}{\eta}(R_{A'})\;,
\end{equation}
where $E\ge0$ is the maximum allowed average energy per mode of the input.
\begin{proof}
The capacity region of the quantum degraded Gaussian broadcast channel is the closure of the union over $n\in\mathbb{N}$ of regions of the form \cite{guha2007classical}
\begin{align}\label{RA}
n\,R_{A'} &\leq \sum_{i\in I} p^{(n)}_i \left(S\left(\hat{\rho}^{A'(n)}_i\right) - \sum_{j\in J}q^{(n)}_j\;S\left(\hat{\rho}^{A'(n)}_{i,j}\right)\right)\;,\\
\label{RB} n\,R_{B'} &\leq S\left(\hat{\rho}^{(n)}_{B'}\right)-\sum_{i\in I}p^{(n)}_i\;S\left(\hat{\rho}^{B'(n)}_i\right)\;,
\end{align}
where $\left\{p^{(n)}_i\,q^{(n)}_j,\;\hat{\rho}^{A(n)}_{i,j}\right\}_{i\in I,\,j\in J}$ is an ensemble of pure encoding states on $n$ copies of the quantum system $A$ and
\begin{align}
\hat{\rho}_A^{(n)} &= \sum_{i\in I,\,j\in J}p^{(n)}_i\,q^{(n)}_j\,\hat{\rho}^{A(n)}_{i,j}\;,\\
\hat{\rho}^{A'B'(n)}_{i,j} &= {\hat{U}_\eta}^{\otimes n}\left(\hat{\rho}^{A(n)}_{i,j}\otimes \left(|0\rangle_B\langle0|\right)^{\otimes n} \right) \hat{U}_\eta^{\dag\otimes n}\;,\\
\hat{\rho}^{A'B'(n)}_i &= \sum_{j\in J} q^{(n)}_j\;\hat{\rho}^{A'B'(n)}_{i,j}\;,\label{rhoBj}\\
\hat{\rho}^{(n)}_{B'} &= \sum_{i\in I} p^{(n)}_i\;\hat{\rho}^{B'(n)}_i\;,
\end{align}
and the average state satisfies the energy constraint
\begin{equation}\label{eq:enc}
\mathrm{Tr}\left[\hat{H}\,\hat{\rho}_A^{(n)}\right] \le n\,E\;.
\end{equation}

Since $S\left(\hat{\rho}^{A'(n)}_{i,j}\right)\ge0$ for any $i\in I$ and $j\in J$, we have from \eqref{RA}
\begin{equation}\label{eq:RA'}
R_{A'} \leq \frac{1}{n}\sum_{i\in I} p^{(n)}_i\,S\left(\hat{\rho}^{A'(n)}_i\right)\;.
\end{equation}
The energy constraint \eqref{eq:enc} implies
\begin{equation}
\mathrm{Tr}\left[\hat{H}\,\hat{\rho}_{B'}^{(n)}\right] \le n\left(1-\eta\right)E\;,
\end{equation}
where $\hat{H}$ is the Hamiltonian on $n$ copies of $B'$, hence
\begin{equation}\label{eq:EB}
S\left(\hat{\rho}_{B'}^{(n)}\right) \le n\,g((1-\eta)E)\;,
\end{equation}
where we have used that quantum thermal Gaussian states maximize the entropy among all the states with the same average energy.
From \eqref{eq:degr} we have for any $i\in I$
\begin{equation}
\hat{\rho}_i^{B'(n)} = \mathcal{E}^{\otimes n}_\frac{1-\eta}{\eta}\left(\hat{\rho}_i^{A'(n)}\right)\;.
\end{equation}
Since the state $\hat{\rho}_A^{(n)}$ has finite average energy, $\hat{\rho}_i^{A'(n)}$ has finite average energy for any $i\in I$, and we have from the hypothesis \eqref{eq:f}
\begin{equation}
S\left(\hat{\rho}_i^{B'(n)}\right) \ge n\,f_\frac{1-\eta}{\eta}\left(\frac{S\left(\hat{\rho}_i^{A'(n)}\right)}{n}\right)\;.
\end{equation}
Since $f_\frac{1-\eta}{\eta}$ is convex and increasing, we have from Jensen's inequality and \eqref{eq:RA'}
\begin{equation}\label{eq:SB}
\frac{1}{n}\sum_{i\in I}p^{(n)}_i\;S\left(\hat{\rho}^{B'(n)}_i\right) \ge f_\frac{1-\eta}{\eta}\left(\frac{1}{n}\sum_{i\in I}p^{(n)}_i\,S\left(\hat{\rho}^{A'(n)}_i\right)\right)\;.
\end{equation}
Putting together \eqref{RB}, \eqref{eq:EB} and \eqref{eq:SB} we get
\begin{equation}\label{eq:RB'}
R_{B'} \le g\left(\left(1-\eta\right)E\right) - f_\frac{1-\eta}{\eta}\left(\frac{1}{n}\sum_{i\in I}p^{(n)}_i\,S\left(\hat{\rho}^{A'(n)}_i\right)\right)\;,
\end{equation}
and the claim follows from \eqref{eq:RA'} and \eqref{eq:RB'}.
\end{proof}
\end{thm}
\begin{cor}[\cite{guha2007classical}]
Assuming Conjecture \ref{conj:MOE} for the quantum-limited attenuator, the capacity region \eqref{eq:R} achievable with the superposition coding is optimal.
\begin{proof}
Follows from Theorem \ref{thm:broadf}.
\end{proof}
\end{cor}
The following Corollary \ref{cor:broadcast} provides the new outer bound to the capacity region of the quantum degraded Gaussian broadcast channel.
We compare in \autoref{fig:broadcast} this outer bound with the previous outer bound provided by the quantum Entropy Power Inequality and with the achievable region \eqref{eq:R}.
\begin{cor}\label{cor:broadcast}
Any achievable rate pair $(R_{A'},\,R_{B'})$ for the quantum degraded Gaussian broadcast channel satisfies \eqref{eq:Rf} with
\begin{equation}\label{eq:flambdaatt}
f_\lambda(x) = g\left(\lambda\,g^{-1}\left(x + g\left(\frac{\lambda}{1-\lambda}\right)\right) + \lambda\right) - g\left(\frac{\lambda}{1-\lambda}\right)\;.
\end{equation}
\begin{proof}
Follows from Theorem \ref{thm:broadf} and Theorem \ref{thm:main}.
\end{proof}
\end{cor}

\section{Bound to the triple trade-off region of the Gaussian quantum-limited attenuator}\label{sec:tradeoff}
We consider the scenario where a quantum channel is used to transmit both classical and quantum information and to generate entanglement shared between the sender and the receiver.
The simplest strategy is the time sharing, which consists in sending only classical information for a fraction of the time, only quantum information for another fraction of the time, and using the channel to generate shared entanglement for the remaining fraction of the time.
The trade-off coding is a more sophisticate strategy that achieves higher rates performing the three tasks simultaneously \cite{hsieh2010entanglement,wilde2012quantumqip,hsieh2010trading}.
A quantum channel where the trade-off coding achieves a remarkable gain with respect to time sharing is the quantum-limited attenuator \cite{wilde2012information}.
Let $C\ge0$, $Q\ge0$ and $G\in\mathbb{R}$ be the rates for classical communication, quantum communication and entanglement generation, respectively, where $G<0$ means that the shared entanglement is consumed instead of being generated.
Then, the quantum-limited attenuator with attenuation parameter $\frac{1}{2}\le\eta\le1$ and input states with maximum average energy per mode $E$ can achieve all the triple of rates $(C,\,Q,\,G)$ such that \cite{wilde2012information,wilde2012quantum}
\begin{align}\label{eq:CQG}
C + 2Q &\le g(\beta\,E) + g(\eta\,E) - g\left(\left(1-\eta\right)\beta\,E\right)\;,\nonumber\\
Q + G &\le g(\eta\,\beta\,E) - g\left(\left(1-\eta\right)\beta\,E\right)\;,\nonumber\\
C + Q + G &\le g(\eta\,E) - g\left(\left(1-\eta\right)\beta\,E\right)
\end{align}
for some $0\le\beta\le1$.
Assuming Conjecture \ref{conj:MOE}, the trade-off region identified by \eqref{eq:CQG} is optimal \cite{wilde2012information,wilde2012quantum}, i.e., any achievable triple of rates $(C,\,Q,\,G)$ satisfies \eqref{eq:CQG}.

We consider also the scenario where a quantum channel is used to transmit both public and private classical information and to generate a secret key shared between the sender and the receiver.
As before the trade-off coding achieves higher rates with respect to the time sharing.
Let $C\ge0$, $P\ge0$ and $K\in\mathbb{R}$ be the rates for public classical communication, private classical communication and key generation, respectively, where $K<0$ means that the shared secret key is consumed instead of being generated.
Then, the quantum-limited attenuator with attenuation parameter $\frac{1}{2}\le\eta\le1$ and input states with maximum average energy per mode $E>0$ can achieve all the triple of rates $(C,\,P,\,K)$ such that \cite{wilde2012information,wilde2012quantum}
\begin{align}\label{eq:CPK}
C + P &\le g(\eta\,E)\;,\nonumber\\
P + K &\le g(\eta\,\beta\,E) - g\left(\left(1-\eta\right)\beta\,E\right)\;,\nonumber\\
C + P + K &\le g(\eta\,E) - g\left(\left(1-\eta\right)\beta\,E\right)
\end{align}
for some $0\le\beta\le1$.
Assuming Conjecture \ref{conj:MOE}, the trade-off region identified by \eqref{eq:CPK} is optimal \cite{wilde2012information,wilde2012quantum}, i.e., any achievable triple of rates $(C,\,P,\,K)$ satisfies \eqref{eq:CPK}.

\begin{figure}
  \includegraphics[width=\columnwidth]{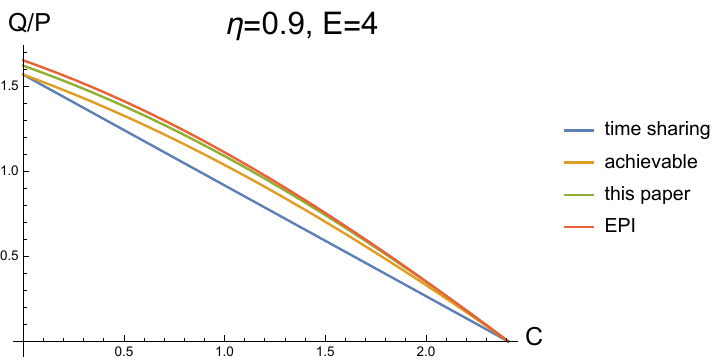}
  \caption{Trade-off region between classical and quantum communication for the Gaussian quantum-limited attenuator with attenuation parameter $\eta=0.9$ and input states with at most $E=4$ average photons per mode.
  The plot compares the regions achievable with time sharing and with the trade-off coding \eqref{eq:CQG} to the outer bounds provided by the quantum Entropy Power Inequality and by Corollary \ref{cor:tradeoff}.
  These regions coincide with the corresponding regions for the trade-off between public and private classical communication, hence the plot encompasses both scenarios.
  All the rates are expressed in nats per channel use.}\label{fig:tradeoff}
\end{figure}

Similarly to the quantum degraded Gaussian broadcast channel, even if Conjecture \ref{conj:MOE} still lacks a proof we can still determine bounds to the triple trade-off regions of the quantum-limited attenuator.
The first of these bounds follows from the quantum Entropy Power Inequality \cite[Appendix C]{qi2016capacities}.
The following Theorem \ref{thm:trf} shows that any lower bound to the output entropy of the multi-mode quantum-limited attenuators in terms of the input entropy implies a bound to their triple trade-off regions.
We then combine Theorem \ref{thm:trf} with Theorem \ref{thm:main} to obtain a new outer bound to the trade-off regions of the quantum-limited attenuator.
\begin{thm}\label{thm:trf}
Let us suppose that for any $n\in\mathbb{N}$, any $0\le\lambda\le1$ and any quantum state $\hat{\rho}$ of an $n$-mode Gaussian quantum system with finite average energy,
\begin{equation}\label{eq:ftr}
S\left(\mathcal{E}^{\otimes n}_\lambda(\hat{\rho})\right) \ge n\,f_\lambda\left(\frac{S(\hat{\rho})}{n}\right)\;,
\end{equation}
where the function $f_\lambda$ is increasing and convex.
Then, any achievable rate triple $(C,\,Q,\,G)$ for the triple trade-off among classical communication, quantum communication and entanglement generation with the Gaussian quantum-limited attenuator with attenuation parameter $\frac{1}{2}\le\eta\le1$ satisfies
\begin{align}\label{eq:CQ1}
C + 2Q &\le g(\eta\,E) + f_\eta^{-1}(g(\beta\,\eta\,E)) - f_\frac{1-\eta}{\eta}(g(\beta\,\eta\,E))\;,\\
\label{eq:QG1}
Q+G &\le g(\beta\,\eta\,E) - f_\frac{1-\eta}{\eta}(g(\beta\,\eta\,E))\;,\\
\label{eq:CQG1}
C + Q + G &\le g(\eta\,E) - f_\frac{1-\eta}{\eta}(g(\beta\,\eta\,E))
\end{align}
for some $0\le\beta\le1$.
Moreover, any achievable rate triple $(C,\,P,\,K)$ for the triple trade-off among public classical communication, private classical communication and key generation satisfies
\begin{align}\label{eq:CP1}
C + P &\le g(\eta\,E)\;,\\
\label{eq:PK1}
P+ K &\le g(\beta\,\eta\,E) - f_\frac{1-\eta}{\eta}(g(\beta\,\eta\,E))\;,\\
\label{eq:CPK1}
C + P + K &\le g(\eta\,E) - f_\frac{1-\eta}{\eta}(g(\beta\,\eta\,E))
\end{align}
for some $0\le\beta\le1$.
\begin{proof}
The set of the achievable triple of rates $(C,\,Q,\,G)$ for the Gaussian quantum-limited attenuator with attenuation parameter $\frac{1}{2}\le\eta\le1$ is the closure of the union over $n\in\mathbb{N}$ of regions of the form \cite{hsieh2010trading,wilde2012quantumqip}
\begin{align}\label{eq:CQ0}
n\left(C + 2Q\right) &\le S\left(\mathcal{E}_\eta^{\otimes n}\left(\hat{\rho}^{(n)}\right)\right)\nonumber\\
&\phantom{\le}+ \sum_{i\in I} p_i^{(n)}\left(S\left(\hat{\rho}^{(n)}_i\right) - S\left(\tilde{\mathcal{E}}_\eta^{\otimes n}\left(\hat{\rho}^{(n)}_i\right)\right)\right)\;,
\end{align}
\begin{align}\label{eq:QG0}
&n\left(Q + G\right)\nonumber\\
&\le \sum_{i\in I} p_i^{(n)}\left(S\left(\mathcal{E}^{\otimes n}_\eta\left(\hat{\rho}^{(n)}_i\right)\right) - S\left(\tilde{\mathcal{E}}^{\otimes n}_\eta\left(\hat{\rho}^{(n)}_i\right)\right)\right)\;,
\end{align}
\begin{align}\label{eq:CQG0}
n\left(C + Q + G\right) &\le S\left(\mathcal{E}_\eta^{\otimes n}\left(\hat{\rho}^{(n)}\right)\right)\nonumber\\
&\phantom{\le} - \sum_{i\in I} p_i^{(n)}\,S\left(\tilde{\mathcal{E}}^{\otimes n}_\eta\left(\hat{\rho}^{(n)}_i\right)\right)\;,
\end{align}
where for any $n\in\mathbb{N}$, $\left\{p^{(n)}_i,\,\hat{\rho}^{(n)}_i\right\}_{i\in I}$ is an ensemble of states of an $n$-mode Gaussian quantum system such that the average state
\begin{equation}
\hat{\rho}^{(n)} = \sum_{i\in I}p^{(n)}_i\,\hat{\rho}^{(n)}_i
\end{equation}
satisfies the energy constraint
\begin{equation}\label{eq:Htr}
\mathrm{Tr}\left[\hat{H}\,\hat{\rho}^{(n)}\right]\le n\,E\;,
\end{equation}
and $\tilde{\mathcal{E}}_\eta$ is the complementary channel of $\mathcal{E}_\eta$.

The energy constraint \eqref{eq:Htr} implies
\begin{equation}
\mathrm{Tr}\left[\hat{H}\,\mathcal{E}^{\otimes n}_\eta\left(\hat{\rho}^{(n)}\right)\right] \le n\,\eta\,E\;,
\end{equation}
and since thermal Gaussian states maximize the entropy among all the states with a given average energy, we have
\begin{equation}\label{eq:SErho}
S\left(\mathcal{E}^{\otimes n}_\eta\left(\hat{\rho}^{(n)}\right)\right) \le n\,g(\eta\,E)\;.
\end{equation}
The concavity of the entropy and \eqref{eq:SErho} imply
\begin{equation}
\sum_{i\in I}p_i^{(n)}\,S\left(\mathcal{E}_\eta^{\otimes n}\left(\hat{\rho}^{(n)}_i\right)\right) \le S\left(\mathcal{E}^{\otimes n}_\eta\left(\hat{\rho}^{(n)}\right)\right) \le n\,g(\eta\,E)\;,
\end{equation}
hence there exists $0\le \beta_n\le 1$ such that
\begin{equation}\label{eq:defbetan}
\sum_{i\in I}p_i^{(n)}\,S\left(\mathcal{E}_\eta^{\otimes n}\left(\hat{\rho}^{(n)}_i\right)\right) = n\,g(\beta_n\,\eta\,E)\;.
\end{equation}
Since the average state $\hat{\rho}^{(n)}$ has finite average energy, $\hat{\rho}_i^{(n)}$ has finite average energy for any $i\in I$.
Since $\tilde{\mathcal{E}}_\eta = \mathcal{E}_\frac{1-\eta}{\eta}\circ \mathcal{E}_\eta$, we have from \eqref{eq:ftr} for any $i\in I$
\begin{equation}\label{eq:fidegr}
S\left(\tilde{\mathcal{E}}_\eta^{\otimes n}\left(\hat{\rho}_i^{(n)}\right)\right) \ge n\,f_\frac{1-\eta}{\eta}\left(\frac{S\left(\mathcal{E}_\eta^{\otimes n}\left(\hat{\rho}_i^{(n)}\right)\right)}{n}\right)\;.
\end{equation}
Since $f_\frac{1-\eta}{\eta}$ is convex, we have from \eqref{eq:fidegr} and Jensen's inequality
\begin{align}\label{eq:fdegr}
&\sum_{i\in I}p_i^{(n)}\,S\left(\tilde{\mathcal{E}}_\eta^{\otimes n}\left(\hat{\rho}_i^{(n)}\right)\right)\nonumber\\
&\ge n\,f_\frac{1-\eta}{\eta}\left(\frac{1}{n}\sum_{i\in I}p_i^{(n)}\,S\left(\mathcal{E}_\eta^{\otimes n}\left(\hat{\rho}^{(n)}_i\right)\right)\right)\nonumber\\
&= n\,f_\frac{1-\eta}{\eta}(g(\beta_n\,\eta\,E))\;,
\end{align}
where in the last step we have used the definition of $\beta_n$.
We have from \eqref{eq:ftr} for any $i\in I$
\begin{equation}
S\left(\mathcal{E}_\eta^{\otimes n}\left(\hat{\rho}^{(n)}_i\right)\right) \ge n\,f_\eta\left(\frac{S\left(\hat{\rho}^{(n)}_i\right)}{n}\right)\;,
\end{equation}
hence
\begin{align}
g(\beta_n\,\eta\,E) &= \frac{1}{n}\sum_{i\in I}p_i^{(n)}\,S\left(\mathcal{E}_\eta^{\otimes n}\left(\hat{\rho}^{(n)}_i\right)\right)\nonumber\\
&\ge f_\eta\left(\frac{1}{n}\sum_{i\in I}p_i^{(n)}\,S\left(\hat{\rho}^{(n)}_i\right)\right)\;,
\end{align}
where we have used Jensen's inequality for $f_\eta$.
Since $f_\eta$ is increasing, we have
\begin{equation}\label{eq:f-1}
\frac{1}{n}\sum_{i\in I}p_i^{(n)}\,S\left(\hat{\rho}^{(n)}_i\right) \le f_\eta^{-1}(g(\beta_n\,\eta\,E))\;.
\end{equation}
The claim \eqref{eq:CQ1} then follows from \eqref{eq:CQ0} together with \eqref{eq:SErho}, \eqref{eq:f-1} and \eqref{eq:fdegr}.
The claim \eqref{eq:QG1} follows from \eqref{eq:QG0} together with \eqref{eq:defbetan} and \eqref{eq:fdegr}.
The claim \eqref{eq:CQG1} follows from \eqref{eq:CQG0} together with \eqref{eq:SErho} and \eqref{eq:fdegr}.

The set of the achievable triple of rates $(C,\,P,\,K)$ is the closure of the union over $n\in\mathbb{N}$ of regions of the form \cite{hsieh2010trading,wilde2012quantumqip}
\begin{align}\label{eq:CP0}
n\left(C + P\right) &\le S\left(\mathcal{E}_\eta^{\otimes n}\left(\hat{\rho}^{(n)}\right)\right)\nonumber\\
&\phantom{\le} - \sum_{i\in I,\,j\in J} p_{i,j}^{(n)}\,S\left(\mathcal{E}_\eta^{\otimes n}\left(\hat{\rho}^{(n)}_{i,j}\right)\right)\;,\\
\end{align}
\begin{align}\label{eq:PK0}
&n\left(P + K\right)\nonumber\\
&\le \sum_{i\in I} p_i^{(n)}\left(S\left(\mathcal{E}^{\otimes n}_\eta\left(\hat{\rho}^{(n)}_i\right)\right) - S\left(\tilde{\mathcal{E}}^{\otimes n}_\eta\left(\hat{\rho}^{(n)}_i\right)\right)\right)\;,
\end{align}
\begin{align}\label{eq:CPK0}
n\left(C + P + K\right) &\le S\left(\mathcal{E}_\eta^{\otimes n}\left(\hat{\rho}^{(n)}\right)\right)\nonumber\\
&\phantom{\le} - \sum_{i\in I} p_i^{(n)}\,S\left(\tilde{\mathcal{E}}^{\otimes n}_\eta\left(\hat{\rho}^{(n)}_i\right)\right)\;,
\end{align}
where for any $n\in\mathbb{N}$, $\left\{p^{(n)}_{i,j},\,\hat{\rho}^{(n)}_{i,j}\right\}_{i\in I,\,j\in J}$ is an ensemble of pure states of an $n$-mode Gaussian quantum system,
\begin{equation}
\hat{\rho}^{(n)}_i = \sum_{j\in J}p^{(n)}_{j|i}\,\hat{\rho}^{(n)}_{i,j}\;,
\end{equation}
and the average state
\begin{equation}
\hat{\rho}^{(n)} = \sum_{i\in I}p^{(n)}_i\,\hat{\rho}^{(n)}_i
\end{equation}
satisfies the energy constraint \eqref{eq:Htr}.

Let $\beta_n$ be as in \eqref{eq:defbetan}.
Then, the claim \eqref{eq:CP1} follows from \eqref{eq:CP0} together with \eqref{eq:SErho} and the property that $S\left(\mathcal{E}_\eta^{\otimes n}\left(\hat{\rho}^{(n)}_{i,j}\right)\right) \ge 0$ for any $i\in I$ and any $j\in J$.
The claim \eqref{eq:PK1} follows from \eqref{eq:PK0} together with \eqref{eq:defbetan} and \eqref{eq:fdegr}.
The claim \eqref{eq:CPK1} follows from \eqref{eq:CPK0} together with \eqref{eq:SErho} and \eqref{eq:fdegr}.
\end{proof}
\end{thm}
\begin{cor}[\cite{wilde2012quantum,wilde2012information}]
Assuming Conjecture \ref{conj:MOE} for the quantum-limited attenuator, the achievable trade-off regions \eqref{eq:CQG} and \eqref{eq:CPK} are optimal.
\begin{proof}
Follows from Theorem \ref{thm:trf}.
\end{proof}
\end{cor}
The following Corollary \ref{cor:tradeoff} provides the new outer bound to the triple trade-off region of the Gaussian quantum-limited attenuator.
In \autoref{fig:tradeoff}, we compare this bound with the previous bound based on the quantum Entropy Power Inequality and with the achievable region \eqref{eq:CQG}.
\begin{cor}\label{cor:tradeoff}
Any achievable rate triple $(C,\,Q,\,G)$ or $(C,\,P,\,K)$ for the trade-off coding with the quantum-limited attenuator satisfies for some $0\le\beta\le1$ \eqref{eq:CQ1}, \eqref{eq:QG1}, \eqref{eq:CQG1} or \eqref{eq:CP1}, \eqref{eq:PK1}, \eqref{eq:CPK1}, respectively, with $f_\lambda$ as in \eqref{eq:flambdaatt}.
\begin{proof}
Follows from Theorem \ref{thm:trf} and Theorem \ref{thm:main}.
\end{proof}
\end{cor}

\section{Conclusions}\label{sec:concl}
We have proven that quantum thermal Gaussian input states minimize the output entropy of the multi-mode quantum Gaussian attenuators and amplifiers that are entanglement breaking and of the quantum Gaussian phase contravariant channels among all the input states with a given entropy (Corollary \ref{cor:EB}).
This result proves the minimum output entropy conjecture (Conjecture \ref{conj:MOE}) for the above channels.
This is the first time that Conjecture \ref{conj:MOE} is proven for multi-mode channels without restrictions on the input states, hence this result significantly extends the cases where the conjecture is known to hold.
We have exploited Corollary \ref{cor:EB} to prove a new lower bound to the output entropy of all the multi-mode quantum Gaussian attenuators and amplifiers (Theorem \ref{thm:main}).
This bound strongly constrains the possible violations of Conjecture \ref{conj:MOE}.
Then, Corollary \ref{cor:EB} and Theorem \ref{thm:main} together provide extremely strong evidence for the general validity of Conjecture \ref{conj:MOE}.

We have applied Theorem \ref{thm:main} to prove new outer bounds to the capacity region of the quantum degraded Gaussian broadcast channel (Corollary \ref{cor:broadcast}) and to the triple trade-off region of the Gaussian quantum-limited attenuator (Corollary \ref{cor:tradeoff}).
The conjectured optimal outer bounds would follow from Conjecture \ref{conj:MOE}, whose proof will be the subject of future work.

\section*{Acknowledgements}
We thank Mark Wilde for useful comments.

\includegraphics[width=0.06\columnwidth]{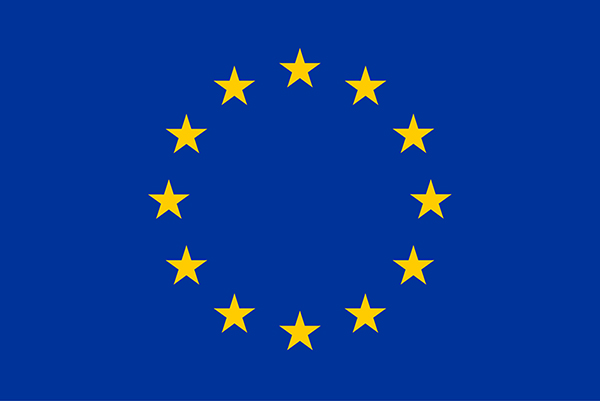}
This project has received funding from the European Union's Horizon 2020 research and innovation programme under the Marie Sk\l odowska-Curie grant agreement No. 792557.

\bibliographystyle{IEEEtran}
\bibliography{biblio}

\begin{IEEEbiographynophoto}{Giacomo De Palma}
was born in Lanciano (CH), Italy, on March 15, 1990.
He received the B.S. degree in Physics and the M.S. degree in Physics from the University of Pisa (Pisa, Italy), in 2011 and 2013, respectively.
He also received the ``Diploma di Licenza'' in Physics and the Ph.D. degree in Physics from Scuola Normale Superiore (Pisa, Italy), in 2014 and 2016, respectively.

He is currently a Marie Sk\l odowska-Curie Individual Fellow at the University of Copenhagen (Copenhagen, Denmark).

His research areas are quantum information and mathematical physics.
He is author of 22 scientific papers published in peer-reviewed journals.
\end{IEEEbiographynophoto}

\end{document}